# Low and Anisotropic Thermal Conductivity in Mixed-Valent $Sn_2S_3$

Xingang Jiang[1], Yongheng Li[1], Weiping Guo[2], Qi Ren[1], Gang Tang[1], Zhong-Zhen Luo[2], Jiawang Hong[1,3*]

[1]*School of Aerospace Engineering, Beijing Institute of Technology, Beijing 100081, China*
[2]*Key Laboratory of Advanced Materials Technologies, International (HongKong Macao and Taiwan) Joint Laboratory on Advanced Materials Technologies, College of Materials Science and Engineering, Fuzhou University, Fuzhou, 350108, P. R. China*
[3]*Beijing Institute of Technology, Zhuhai Beijing Institute of Technology (BIT)*

## Abstract

Compounds of Sn, such as SnSe and SnS, exhibit novel phonon characteristics and low thermal conductivity, making them emerging star materials in the thermoelectric family. In this work, through the Boltzmann transport equation scheme and the Wigner thermal transport model, quasi-1D mixed-valent $Sn_2S_3$ were found to exhibit a low thermal conductivity along *c*-axis with a weak temperature dependence. The low thermal conductivity is attributed to the anharmonic rattling vibrations of weakly bonded Sn(II) atoms, which are influenced by the coulomb interaction of lone pairs at adjacent Sn(II) atoms. The rattling of Sn(II) induces low-frequency flat optical phonons and avoids crossing behavior. The atomic displacements and mean square displacement (MSD) analysis reveal that Sn(II) atoms exhibit significantly greater and anisotropic displacements compared to Sn(IV) and S, confirming that Sn(II) behaves as a rattler. The results obtained from this work suggest an opportunity to discover low thermal conductivity in mixed-valent compounds.

## Introduction

Materials with low thermal conductivity have a wide range of practical applications, such as thermoelectric energy conversion[1] and thermal management[2]. The efficiency of thermoelectric materials is determined by the thermoelectric figure of merit, $ZT = S^2\sigma T/(\kappa_e+\kappa_l)$, where $S$ is the Seebeck coefficient, $\sigma$ is the electrical conductivity, $\kappa_e$ is the electronic thermal conductivity, $\kappa_l$ is the lattice thermal conductivity, and $T$ is the absolute temperature[3,4]. Enhancing thermoelectric performance typically involves balancing two key factors[5]: increasing electrical performance and reducing thermal conductivity. However, the electronic contribution to the thermal conductivity ($\kappa_e$) is strongly coupled with electrical conductivity and the Seebeck coefficient, making the optimization of thermoelectric performance challenging. In contrast, heat transfer by lattice vibration ($\kappa_l$) is a relatively independent factor affecting thermoelectric properties. Therefore, searching for materials with intrinsically low $\kappa_l$ and understanding their thermal transport physics is crucial for exploring novel thermoelectric materials[6-14].

Phonon interactions are closely related to the properties and strength of crystal bonding[15-17]. Recently, the concepts of rattling[18-20] and lone pairs[21-23] have been identified as effective strategies for achieving low thermal conductivity in semiconductors while maintaining the good electronic transport. Generally, atomic rattling leads to the weaker chemical bonds within the crystal structure due to weak atomic interactions. Introducing rattling atoms which are characterized by large amplitude vibrations, has emerged as one of the most significant phenomena for effectively reducing the thermal conductivity in cage materials such as clathrates and skutterudites[24]. Furthermore, the incorporation of low-frequency localized vibrations by rattler atoms within rigid lattices markedly diminishes lattice thermal conductivity. This phenomenon arises due to the weak coupling of these localized vibrations to the lattice framework, similar to the behavior exhibited by rattler atoms[19]. These localized vibrations strongly scatter phonons via rattling scattering mechanisms, a phenomenon also observed in compounds containing Cu(I)[25] and Ag(I)[26-29] atoms. Certain structural configurations exhibit weak atomic interactions, allowing weakly bonded atoms to be approximated as rattlers[30,31]. Additionally, materials with lone pairs can display rattling behavior. The electrostatic repulsion force exerted by these lone pairs leads to atomic rattling, resulting in strong anharmonicity and low phonon group velocities[21,32,33]. Mixed-valent materials constitute a rare and intriguing class of compounds, often exhibiting rattling atoms[18]. This phenomenon primarily stems from the unique spatially active lone pair electrons, which create anisotropic forces on atoms and give rise to the characteristic rattling motion. $Sn_2S_3$ serves as a typical example of such materials. However, an in-depth understanding of how rattling atoms influence thermal conductivity of mixed-valent $Sn_2S_3$ is still far from complete.

The Sn-S family of materials, including SnS, $SnS_2$, and $Sn_2S_3$[34], has garnered attention as promising thermoelectric materials[35-37]. These materials composed of abundant and non-toxic elements and exhibit excellent electrical properties, making them suitable for applications in field-effect transistors, photodetectors and electroluminescent devices[38]. Among them, $Sn_2S_3$ stands out with its quasi-1D structure and pronounced anisotropy, offering intriguing opportunities for thermal management in devices. Moreover, $Sn_2S_3$ is a mixed-valent compound featuring two different oxidation states of Sn: Sn(II) and Sn(IV), resulting in a noticeable bonding hierarchy within its structural sublattices[39], where Sn(II) atoms with lone pairs[40] show very weak and anisotropic bonding with their neighboring S atoms, potentially leading to rattling behavior. The low thermal conductivity of $Sn_2S_3$ has been experimentally measured in the previous work[41], and its thermoelectric properties have been explored computationally[40]. Nevertheless, the rattling mechanism and the influence on the low thermal conductivity remain ambiguous and significant effort is still needed.

In this work, we focused on studying lattice dynamics and phonon transport in $Sn_2S_3$. Density functional theory (DFT) calculations results indicate that $Sn_2S_3$ exhibits low thermal conductivity. This phenomenon is attributed to the large atomic displacements of Sn(II), which resulting in rattling-like thermal damping. Moreover, due to the quasi-1D structure, the thermal conductivity of $Sn_2S_3$ is highly anisotropic. Contrary to the long-standing view that acoustic phonons with long mean-free paths invariably are the primary heat carriers contributing to $\kappa_l$, we show that 63% of $\kappa_l$ of $Sn_2S_3$ is contributed by optical phonons. These findings have profound implications for thermoelectric applications, particularly in systems with anisotropic and weakly bonding.

# Methods
## Computational details

In this work, the first-principles calculations were performed using the Vienna Ab initio Simulation Package (VASP) software[42,43]. The electron-ion interactions were described using the Perdew-Burke-Ernzerhof (PBE) functional within the generalized gradient approximation (GGA)[44]. The projector augmented plane wave (PAW) pseudopotentials were adopted to evaluate the interaction between electrons and ions, and the Grimme's DFT-D1 approach[45] was considered in all the calculations since $Sn_2S_3$ were quasi-1D van der Waals structure. The plane wave energy cutoff for all calculations was set to 550 eV. the convergence criteria for Hellmann-Feynman force and total energy were $10^{-3}$ eV/Å and $10^{-6}$ eV/cell.

The ab initio molecular dynamics (AIMD) simulations were performed at multiple temperatures ranging from 200 K to 700 K (in increments of 100 K) using the Born-Oppenheimer molecular dynamics framework, with temperature control implemented via the Nose thermostat under the NVT ensemble. A simulation time of 3 ps with a time step of 2 fs, and a 2×4×2 supercell of 320 atoms were used. The temperature dependent effective potentials (TDEP) software[46,47] was used to preprocess the AIMD simulation data, and the temperature-dependent harmonic, cubic and quartic interatomic force constants (IFCs) were extracted using the hiphive package[48]. The thermal conductivity was then calculated from these IFCs with the ShengBTE and Fourphonon package[49,50]. The convergence with respect to the reciprocal space grid was tested, and finally a 11×26×7 q-point grid was used to obtain the thermal conductivity.

In the framework of the Wigner transport equation[51,52], the lattice thermal conductivity including the particle-like and wave-like contributions:

$$\kappa_l^{\alpha\beta} = \kappa_p^{\alpha\beta} + \kappa_c^{\alpha\beta} \tag{1}$$

particle-like contribution ($\kappa_p^{\alpha\beta}$) to lattice thermal conductivity resulting from the diagonal ($s = s'$) terms of the Wigner heat-flux operator:

$$\kappa_p^{\alpha\beta} = \frac{1}{VN_q} \sum_{qs} C_{q,s} v_{q,s}^\alpha v_{q,s}^\beta \tau_{q,s} \tag{2}$$

The computation of wavelike contributions described by off-diagonal ($s \neq s'$) terms, describes the tunneling between phonon branches $s$ and $s'$:

$$\kappa_c^{\alpha\beta} = \frac{\hbar^2}{k_B T^2 VN} \sum_q \sum_{s \neq s'} \frac{\omega_{q,s} + \omega_{q,s'}}{2} v_{q,ss'}^\alpha \cdot v_{q,ss'}^{s,s'}$$

$$\times \frac{\omega_{q,s} n_{q,s}(n_{q,s}+1) + \omega_{q,s'} n_{q,s'}(n_{q,s'}+1)}{4(\omega_{q,s} - \omega_{q,s'})^2 + (\Gamma_q^s + \Gamma_q^{s'})^2} \times (\Gamma_q^s + \Gamma_q^{s'}) \tag{3}$$

where $\alpha$ and $\beta$ are indexing the cartesian indices, $V$ and are the $N_q$ volume of the unit cell and the number of sampled phonons in the first Brillouin zone. $C_q^s, v_q^s, \tau_q^s, \omega_q^s, v_q^{s,s'}$ and $\Gamma_q^s$ are the heat capacity, group velocity, lifetime, phonon frequency, velocity operator, and scattering rate of a phonon mode, respectively. $n_q^s=1/[e^{\left(\frac{\hbar\omega_{q,s}}{k_BT}\right)}-1]$ is the Bose-Einstein distribution.

We used the PHONOPY package[53] to obtain the phonon dispersion and phonon DOS of $Sn_2S_3$ with a supercell of 2×4×2, containing 320 atoms, on the basis of density functional perturbation theory (DFPT). The atomic displacement parameters (ADPs) were calculated from the second-order forces. In addition, the projected crystal orbital Hamilton population (pCOHP)[54] was used to evaluate the bonding variation and interaction in the $Sn_2S_3$ with the help of Local Orbital Basis Suite Towards Electronic-Structure Reconstruction (LOBSTER).

**Results**

As illustrated in **Fig. 1a**, the crystal structure of $Sn_2S_3$ exhibits quasi-1D characteristics, with bonding primarily along the *b*-axis and van der Waals interactions along the *a*- and *c*-axis, where visible van der Waals gaps exist. The optimized lattice constants for the conventional cell of $Sn_2S_3$ with space group *Pnma* are *a*=8.84 Å, *b*=3.78 Å, *c*=13.98 Å, very close to experiment values[55] and other computational data[56]. The unit cell of $Sn_2S_3$ contains 20 atoms, with the two oxidation states of Sn atoms represented by cyan and purple colors, and the Sn(II)-S bond is longer than the Sn(IV)-S bond. This quasi-1D structure implies significant anisotropic thermal transport in $Sn_2S_3$.

To further understand the bonding formability with respect to the variation of orbital states of atoms, a detailed analysis on -pCOHP is carried out. A positive -pCOHP value indicates bonding interactions, while a negative value indicates the anti-bonding interactions. Generally, the active lone pairs are dominantly contributed by the s-orbital and exhibit anti-bonding interactions. This suggests that the electrostatic force exerted by lone pair electrons may weaken the Sn(II)-S bond strength, making Sn(II) more prone to rattling, and has been confirmed through COHP analysis as shown in **Fig. 1b**. Additionally, we plotted the potential energy barriers for atomic displacements as a function of atomic displacement around the equilibrium positions along the *c*-axis, revealing that Sn(II) has the lowest energy barrier as shown in **Fig. 1c** and **Fig. S1**, indicating the weakest bonding with the lattice and the smallest linear restoring force. Similar behavior is observed in other systems with lone pairs such as InTe and $TlInTe_2$. In(I) and Tl(I), which possess lone pairs, exhibit the flattest energy potential wells and have the largest root-mean-square displacements[32,57].

Molecular dynamics serves as a powerful tool for investigating atomic motion behavior. To obtain more direct evidence of the rattling behavior of Sn(II) atoms, we conducted a Born-Oppenheimer molecule dynamic simulations of 50000 steps with a time step of 2 fs simulation based on the NVT ensemble within a 2×3×2 supercell containing 160 atoms to capture the trajectories of individual atoms at a temperature of 300 K. The resulting plots depict the trajectories of Sn(IV), Sn(II), and S, as illustrated in the **Fig. S2**. Notably, our analysis reveals that the atomic displacements of Sn(IV) and S exhibit isotropic behavior, whereas those of Sn(II) is anisotropic, which attributed to the electrostatic repulsion force of the lone pairs. Specifically, the displacement of Sn(II) atoms along the *c*-axis significantly exceeds that observed in the *a*- and *b*-axis. This may because Sn(II) have large distance and weakly bonding from the adjacent atoms, which provides enough space for vibration freely with larger displacement. Furthermore, mean square displacement (MSD) calculations in **Fig. S3.** indicate that, within the temperature range of 300 K to 1000 K, Sn(II) experiences substantially greater displacement compared to Sn(IV) and S. This pronounced difference suggests that Sn(II) behaves as a rattler.

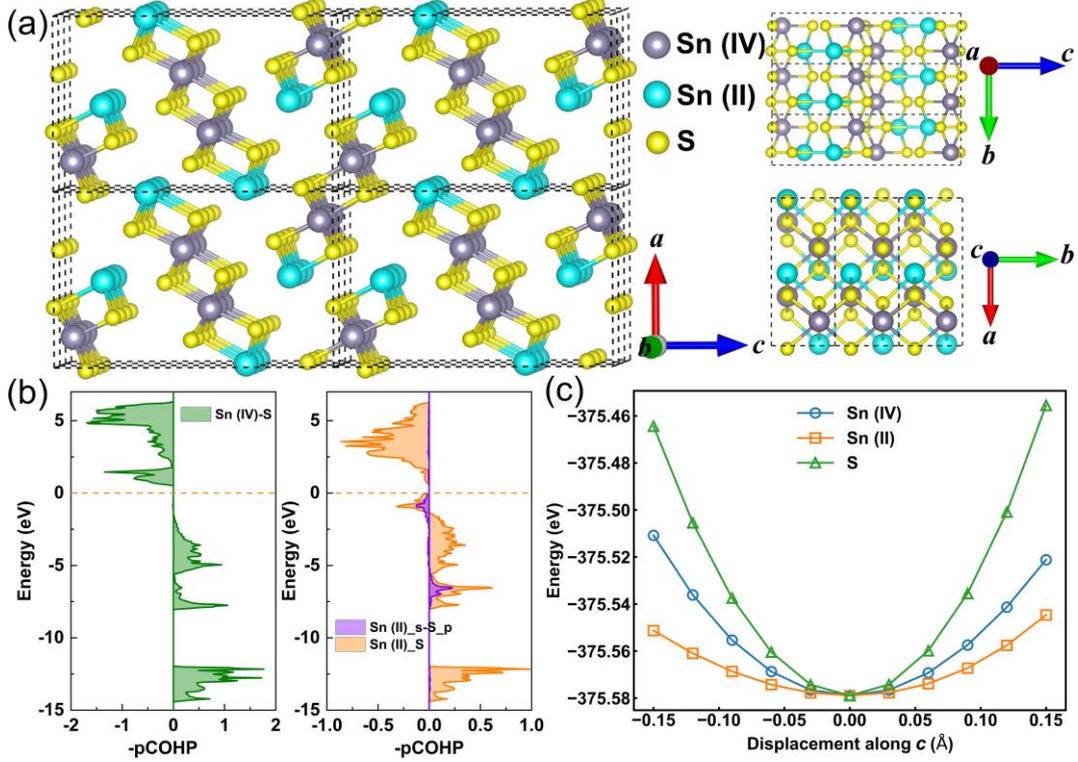

**Fig. 1.** (a) Lattice structures of $Sn_2S_3$ as viewed against the crystalline *b*-, *a*-, and *c*-axis. (b) XRD patterns for $Sn_2S_3$ compounds, insert is the crystal sample of $Sn_2S_3$. (c) potential energy *vs* displacements along crystallographic *c*-axis of unit cell.

The DFT-calculated phonon dispersion relation is shown in **Fig. 2a,** it has been observed that $Sn_2S_3$ exhibits very flat low frequency optical branches, and intersects with the acoustic branches throughout the Brillion zone, this strong interaction results in enhanced phonon scattering that reduces the lattice thermal conductivity, and such a phenomenon also reported in other rattling systems. The optical phonons are relatively flat in the Γ-X and Γ-Z directions and intertwined with acoustic phonons, which is a typical characteristic of rattling systems. Whereas they exhibit a significant slope in the Γ-Y direction, implying that the thermal contribution from the optical branches may not be negligible. Additionally, an avoid-crossing phenomenon was identified in the T-Z direction. Another significant point to be noticed is that $Sn_2S_3$ exhibits phonon softening along the high symmetry point Z-U, this is favorable for achieving low $\kappa_l$ [58,59]. The calculated projected phonon densities of states (PDOS) are shown in **Fig. 2b.** The S atoms predominantly contribute to the high-energy (>22 meV) optical modes due to their lighter atomic mass, whereas the Sn atoms dominate the acoustic phonons as well as all the low-energy optical phonons (<22 meV) because of their large stoichiometric ratio and heavier atomic mass. Importantly, the strong coupling of low frequency phonon branches produces a sharp peak in the phonon density of stats, which a typical indicator of rattling[19,27,30], indicating that the lattice thermal conductivity may be largely reduced.

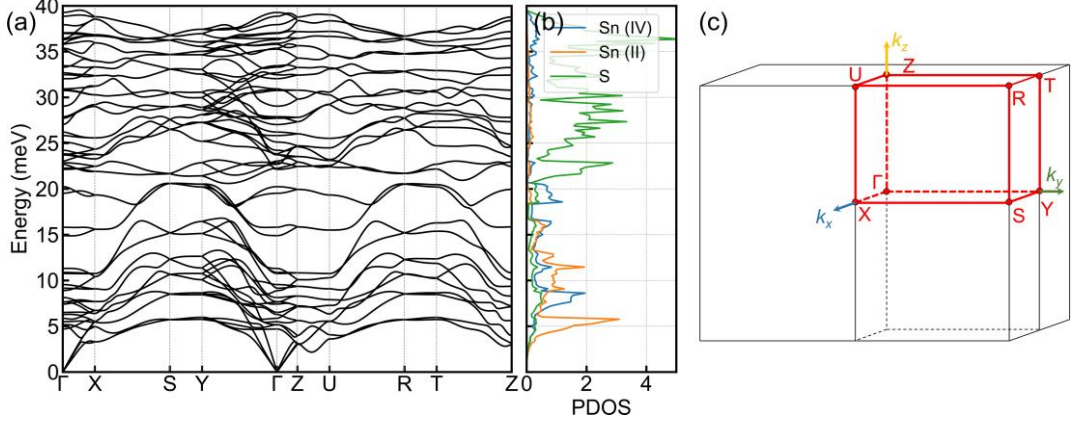

**Fig. 2.** (a) Calculated phonon dispersion and (b) PDOS for $Sn_2S_3$, (c) and the first Brillouin zone of $Sn_2S_3$.

The normalized trace of interatomic force constants (IFC) was used to evaluate the bonding strength of $Sn_2S_3$, in which as follows[15]:

$$normalized\ trace\ of\ IFC = \frac{\frac{\partial^2 E}{\partial R_{0,x} \partial R_{n,x}} + \frac{\partial^2 E}{\partial R_{0,y} \partial R_{n,y}} + \frac{\partial^2 E}{\partial R_{0,z} \partial R_{n,z}}}{\frac{\partial^2 E}{\partial R_{0,x} \partial R_{0,x}} + \frac{\partial^2 E}{\partial R_{0,y} \partial R_{0,y}} + \frac{\partial^2 E}{\partial R_{0,z} \partial R_{0,z}}} \quad (4)$$

In the above expression, $\frac{\partial^2 E}{\partial R_{0,x} \partial R_{n,x}}$ represents the second-order force constant along x-direction between the origin atom (described as "0" in $R_{0,x}$) and the n-th neighbor atom (described as "n" in $R_{n,x}$). The $\frac{\partial^2 E}{\partial R_{0,x} \partial R_{0,x}}$ represents the self-interaction force constant along *x*-direction, which means the force constant of one specific atom when the atom itself is displaced. The bond length between Sn(II) and the six surrounding S atoms ranges from 2.67 to 3.36, while the bond length between Sn(IV) and the six surrounding S atoms is in the range of 2.51 to 2.62. The range of normalized trace of IFC for Sn(II) and the six surrounding S atoms is from 0.0024 to 0.3008, whereas the range of normalized trace of IFC for Sn(IV) and the six surrounding S atoms is range of 0.1345 to 0.2027, as shown in **Fig. S4**. Compared with Sn(IV), Sn(II) exhibits greater anisotropy in bonding with the surrounding S atoms and is therefore more likely to show rattling behavior. More importantly, the total average frequency of $Sn_2S_3$ is denoted as 22.57 meV, further examination of the average frequency for each atom reveals that Sn(II) exhibits an average frequency of 11.21 meV, while S has an average frequency of 28.91 meV. Notably, the average frequency of Sn(II) is significantly lower than the overall average, whereas the average frequency of S exceeds the average value. Specifically, Sn(II) vibrates at approximately 38.78% of the frequency observed for S, indicating characteristic rattling behavior associated with Sn(II). This distribution of average phonon energy of various atoms in $Sn_2S_3$ is similar to some structures with rattling characteristics[9,30].

The 3,4ph and 3ph calculated thermal conductivity values along *c*-axis at 300 K were 1.34 and 1.48 W m$^{-1}$ K$^{-1}$, and the temperature dependence of the lattice thermal conductivity follows $T^{-1.03}$ and $T^{-1.04}$, respectively. Phonon propagation in crystals is generally treated as particle-like, as described by Peierls' formulation of the BTE. In recent works[8], the non-traditional wave-like diffusion of phonons plays important roles in interpretation lattice thermal conductivity in solid materials, the two conduction mechanisms are unified in the Wigner transport equation as discussed in method, by using the 3,4ph+off-diagonal (OD) model, the lattice thermal conductivity exhibits a weak temperature dependence of $T^{-0.68}$. This behavior deviates from the $T^{-1}$ trend predicted by the conventional phonon gas model. **Fig. 3c** illustrates that the lattice thermal conductivity is predominantly contributed by particle-like phonons at 300 K, while the wave-like thermal conductivity primarily arises from optical phonons. The wave-like thermal conductivity is approximately 0.27 W m$^{-1}$ K$^{-1}$, significantly lower than that of the particle-like thermal conductivity. In **Fig. 3d**, the density of states for wave-like thermal conductivity is presented, characterized by the frequencies of various phonon pairs at 300 K, labeled as phonon Energy1 and phonon Energy2. This indicates that the contributions to the wave-like channel mainly originate from the high-frequency region.

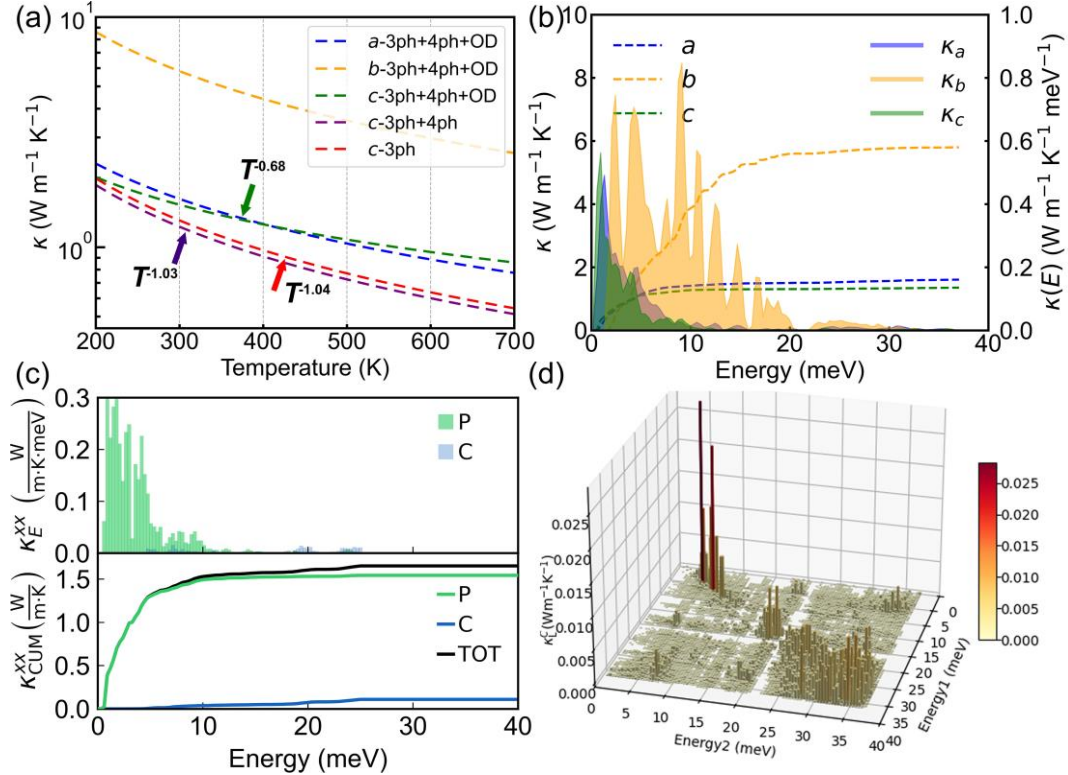

**Fig. 3.** (a) Calculated lattice thermal conductivity $\kappa$ as a function of temperature in $Sn_2S_3$. (b)The anisotropic $\kappa_l$ are distinguished by the different colored lines, namely blue, green, and orange, corresponding one-to-one to the colors in Fig. 1a, the shaded regions indicate the spectral $\kappa$ as a function of energy. (c)The energy dependent thermal conductivity density of states and cumulative thermal conductivity from particle-like and wave-like contributions along *c*-axis (green line in Fig. 1a) at 300 K, correspond to Fig (d) Three-dimensional visualization model $\kappa_l^C$ of the contributions to the coherences thermal conductivity calculated by the 3ph+4ph+OD model along the *c* axis at 300 K.

Quasi-one-dimensional materials usually exhibit significant anisotropic thermal conductivity. The calculated lattice thermal conductivity based on 3ph+4ph+OD model of $Sn_2S_3$ along the *a*- and *c*-axis is 1.73 and 1.59 W m$^{-1}$ K$^{-1}$, respectively. Notably, along the *b*-axis, the thermal conductivity is significantly higher, reaching 6.55 W m$^{-1}$ K$^{-1}$. This exhibit pronounced anisotropy, with an anisotropy ratio of 3.86 at room temperature, indicating a substantial difference in heat conduction along different axis. This is attributed to the fact that the *a*- and *c*-axis correspond to van der Waals directions, while the *b*-axis corresponds to the bonding direction. Phonon group velocity analysis reveals that the group velocity of acoustic phonons is greater than that of optical phonons (as shown in **Fig. 4a**). Another key factor influencing thermal conductivity is the phonon lifetime. As phonon lifetimes plotted in **Fig. 4b**, the phonon lifetimes of acoustic phonons are also longer than those of optical phonons. This indicates that acoustic phonons are likely the primary contributors to the thermal conductivity of $Sn_2S_3$. Further cumulative thermal conductivity analysis shows that the thermal conductivity along the *a*- and *c*-axis is predominantly contributed by acoustic phonons. Interestingly, approximately 63% of the thermal conductivity along the *b*-axis is contributed by optical phonons, which contradicts conventional understanding.

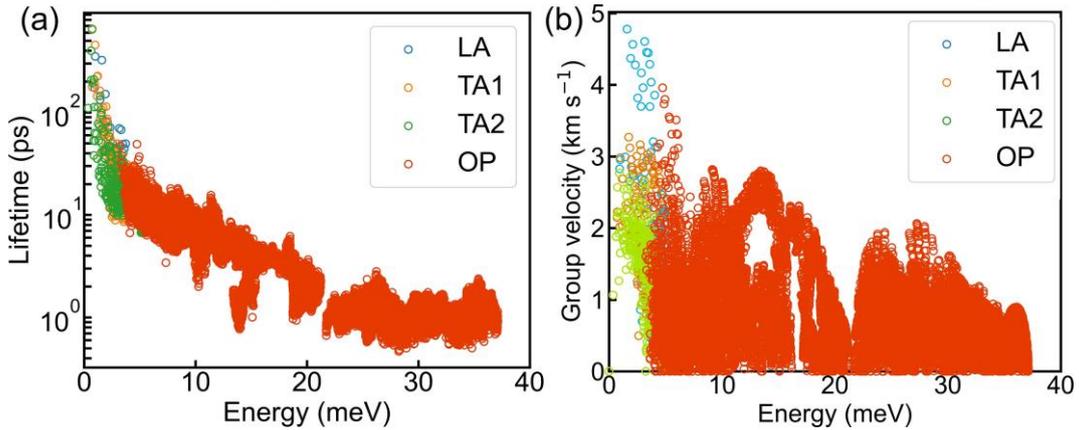

**Fig. 4.** Energy dependence of (a) lifetime (computational details can be found in the supplementary information) and (b) group velocity.

To investigate the anomalous large ratio of contribution of optical phonons to thermal conductivity, we performed a projection analysis of the group velocity energy density calculated by[60]

$$\frac{(v^\mu)^2}{|v|}(\omega) = \left(\frac{\Delta q}{2\pi}\right)^3 \sum_{\alpha,q} \frac{(v^\mu_{\alpha,q})^2}{|v^\mu_{\alpha,q}|} \delta(\omega - \omega_{\alpha,q}) / \sum_\alpha g_\alpha(\omega) \tag{5}$$

where $\alpha$ sums over all modes, q sums over all lattice vectors in the Brillouin zone, $v^\mu_{\alpha,q}$ is the group velocity along the $\mu$ direction and $g_\alpha(\omega)$ is phonon density of states, as shown in **Fig. 5a**. The results indicate that in the range of 5 meV to 32 meV, the phonon group velocity along b-axis is significantly higher than in the a- and c-axis. Additionally, we observed that below 5 meV, the phonon group velocity in the c-axis is higher than in the a-axis, these results are consistent with the quasi-1D atomic structure of Sn$_2$S$_3$, which has the strongest bonding along b-axis. Consequently, we further calculated the mode-dependent phonon group velocities as shown in **Fig. 5b**. It was found that while the group velocities of acoustic phonons are similar, the group velocities of the leading two optical modes along Y-Γ are considerably higher than those along Γ-X and Γ-Z. This is related to the coupling between low-frequency optical phonons and acoustic phonons.

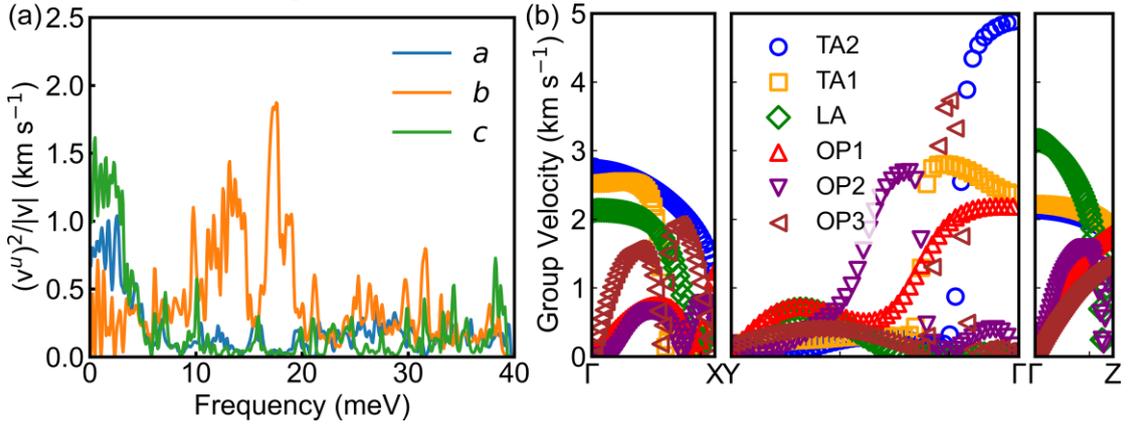

**Fig. 5.** (a) Average group velocity of phonon modes along the a-, b- and c-axis, respectively, showing high anisotropy for optical phonons between 5 and 32 meV. (b) mode-resolved group velocity of the LA, TA1, TA2 mode acoustic phonons and the optical phonon modes at room temperature.

In order to elucidate the underlying mechanism of the thermal properties in mixed-valent materials, it is crucial to consider their unique behaviors due to specific electronic configurations and bonding environments. The weak Sn(II)-S bonds and the lone pairs could strength crystal anharmonicity and further scatter phonons, resulting in low lattice thermal conductivity in Sn$_2$S$_3$. The strength of anharmonicity can be estimated by the Grüneisen parameter, which characterizes the relationship between phonon frequency and volume change, as defined: $\gamma_i = -\frac{V}{\omega_i}\frac{\partial \omega_i}{\partial V}$, where V is the volume of unit cell and $\omega_i$ is the frequency of the $i$th phonon branch. Therefore, we computed the mode Gruneisen parameters projected onto the phonon spectrum for Sn$_2$S$_3$. As shown in **Fig. 6a**, the low-frequency optical branches and acoustic branches exhibit the largest Gruneisen parameters. From the phonon density of states, it is evident that this behavior corresponds to Sn(II). The significant anharmonicity is further evidenced by the frozen potential calculations of the third optical mode at Γ (Γ-OP-3) mode, as shown in **Fig. 6b**. The harmonic fittings, or quadratic fits, of the frozen phonon potential deviate substantially from the DFT-calculated potential energy surface. This deviation indicates a nonlinear dependence of force on atomic displacement, highlighting the strong anharmonicity in these phonon modes.

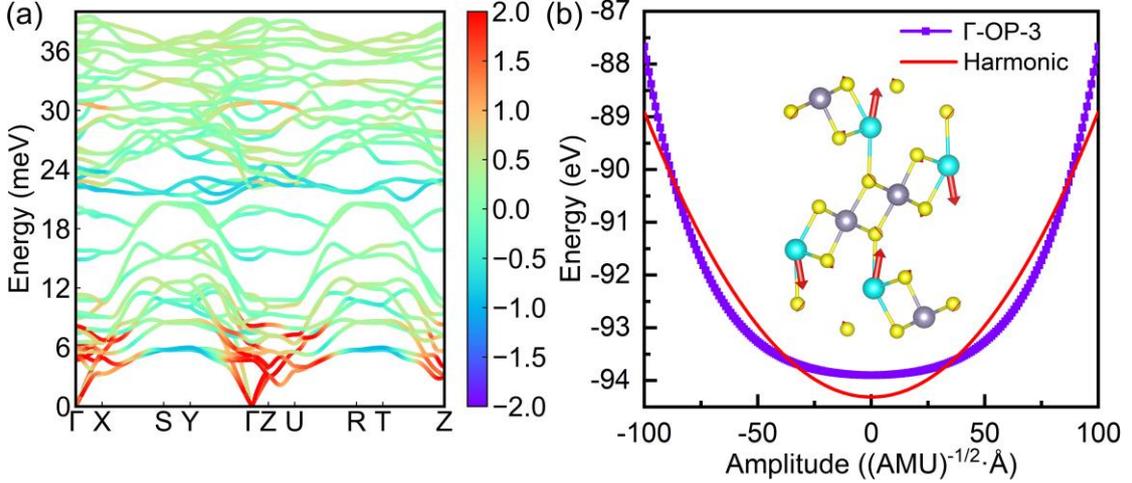

**Fig. 6.** (a) Projection of Grüneisen Parameters onto the Phonon dispersion, the color bar represents the values of Grüneisen Parameters. (b) Frozen phonon potential of Γ-OP-3, where AMU is atomic mass unit, dots are frozen potential calculations of the Γ-OP-3 mode and the red line is harmonic fittings.

In summary, the origin of the intrinsically low and weak temperature dependent lattice thermal conductivity of $Sn_2S_3$ is studied in detail. Rattling of Sn(II) atoms induced by lone pairs is found to be responsible for the low lattice thermal conductivity. Their vibrations cause low-frequency optic phonons and further contribute to the rattling-like thermal damping and reduced lattice thermal conductivity. The anisotropy ratio of heat transfer of $Sn_2S_3$ has reached 3.86, since the one-dimensional channel formed between the chains of $Sn_2S_3$ significantly hinders the heat carrying of acoustic phonons and thereby obstructs the heat conduction. Anomalously large thermal transport contribution in the *b*-axis of the optical branch is due to the large phonon group velocity of the optical phonons in the frequency range from 5 meV to 32 meV. The rattling of Sn(II) atoms in the typical mixed-valent compound $Sn_2S_3$, due to the anisotropic bonding strength, provides unique insights for the search for low thermal conductivity materials, especially in other mixed-valent compounds.

**Acknowledgements**
The work at Beijing Institute of Technology is supported by the National Key R&D Program of China (2021YFA1400300), BIT Research and Innovation Promoting Project (Grant No. 2023YCXZ002), National Natural Science Foundation of China (52102218), and the Fujian Science & Technology Innovation Laboratory for Optoelectronic Information of China (2021ZZ127).

# Supporting Information

## Low and Anisotropic Thermal Conductivity in Mixed-Valent Sn$_2$S$_3$

Xingang Jiang[1], Yongheng Li[1], Weiping Guo[2], Qi Ren[1], Zhong-Zhen Luo[2], Jiawang Hong[1,3*]

[1]School of Aerospace Engineering, Beijing Institute of Technology, Beijing 100081, China
[2]Key Laboratory of Advanced Materials Technologies, International (HongKong Macao and Taiwan) Joint Laboratory on Advanced Materials Technologies, College of Materials Science and Engineering, Fuzhou University, Fuzhou, 350108, P. R. China
[3]Beijing Institute of Technology, Zhuhai Beijing Institute of Technology (BIT)

In this work, we consider the anharmonic phonon interactions up to the four-phonon scattering process. In addition, we also take into account the isotope effect.

$$\frac{1}{\tau_\lambda^0} = \frac{1}{N}\left(\sum_{\lambda'\lambda''}^{(+)} \Gamma^{(+)}_{\lambda\lambda'\lambda''} + \frac{1}{2}\sum_{\lambda'\lambda''}^{(-)} \Gamma^{(-)}_{\lambda\lambda'\lambda''}\right) + \frac{1}{N}\sum_{\lambda'}^{(iso)} \Gamma^{(iso)}_{\lambda\lambda'}$$
$$+ \frac{1}{N}\left(\sum_{\lambda'\lambda''\lambda'''}^{(++)} \frac{1}{2}\Gamma^{(++)}_{\lambda\lambda'\lambda''\lambda'''} + \sum_{\lambda'\lambda''\lambda'''}^{(+-)} \frac{1}{2}\Gamma^{(+-)}_{\lambda\lambda'\lambda''\lambda'''} + \sum_{\lambda'\lambda''\lambda'''}^{(--)} \frac{1}{6}\Gamma^{(--)}_{\lambda\lambda'\lambda''\lambda'''}\right) \quad (S1)$$

where N is the total grid of q points. The superscripts ($\pm$) or ($\pm\pm$) of $\Gamma$ represent the scattering rates for three- and four-phonon process, given by:

$$\Gamma^{(+)}_{\lambda\lambda'\lambda''} = \frac{\hbar\pi}{4}\frac{n^0_{\lambda'} - n^0_{\lambda''}}{\omega_\lambda \omega_{\lambda'} \omega_{\lambda''}}\left|V^{(+)}_{\lambda\lambda'\lambda''}\right|^2 \delta(\omega_\lambda + \omega_{\lambda'} - \omega_{\lambda''}) \quad (S2)$$

$$\Gamma^{(-)}_{\lambda\lambda'\lambda''} = \frac{\hbar\pi}{4}\frac{n^0_{\lambda'} + n^0_{\lambda''} + 1}{\omega_\lambda \omega_{\lambda'} \omega_{\lambda''}}\left|V^{(-)}_{\lambda\lambda'\lambda''}\right|^2 \delta(\omega_\lambda - \omega_{\lambda'} - \omega_{\lambda''}) \quad (S3)$$

$$\Gamma^{(++)}_{\lambda\lambda'\lambda''\lambda'''} = \frac{\hbar^2\pi}{8N}\frac{(1+n^0_{\lambda'})(1+n^0_{\lambda''})n^0_{\lambda'''}}{n^0_\lambda}\left|V^{(++)}_{\lambda\lambda'\lambda''\lambda'''}\right|^2 \frac{\delta(\omega_\lambda + \omega_{\lambda'} + \omega_{\lambda''} - \omega_{\lambda'''})}{\omega_\lambda \omega_{\lambda'} \omega_{\lambda''} \omega_{\lambda'''}} \quad (S4)$$

$$\Gamma^{(+-)}_{\lambda\lambda'\lambda''\lambda'''} = \frac{\hbar^2\pi}{8N}\frac{(1+n^0_{\lambda'})n^0_{\lambda''}n^0_{\lambda'''}}{n^0_\lambda}\left|V^{(+-)}_{\lambda\lambda'\lambda''\lambda'''}\right|^2 \frac{\delta(\omega_\lambda + \omega_{\lambda'} - \omega_{\lambda''} - \omega_{\lambda'''})}{\omega_\lambda \omega_{\lambda'} \omega_{\lambda''} \omega_{\lambda'''}} \quad (S5)$$

$$\Gamma^{(--)}_{\lambda\lambda'\lambda''\lambda'''} = \frac{\hbar^2\pi}{8N}\frac{n^0_{\lambda'}n^0_{\lambda''}n^0_{\lambda'''}}{n^0_\lambda}\left|V^{(--)}_{\lambda\lambda'\lambda''\lambda'''}\right|^2 \frac{\delta(\omega_\lambda - \omega_{\lambda'} - \omega_{\lambda''} - \omega_{\lambda'''})}{\omega_\lambda \omega_{\lambda'} \omega_{\lambda''} \omega_{\lambda'''}} \quad (S6)$$

Where $n^0_\lambda$ being the phonon Bose-Einstein distribution at equilibrium, $\omega_\lambda$ being the phonon frequency for a certain mode $\lambda$. Conservation of energy is enforced by the Dirac delta function $\delta$. The matrix elements V are given by the Fourier transformation of force constants, or transition probability matrices:

$$V_{\lambda\lambda'\lambda''}^{(\pm)} = \sum_{ijk}\sum_{\alpha\beta\gamma} \Phi_{ijk}^{\alpha\beta\gamma} \frac{e_\alpha^\lambda(i) e_\beta^{\pm\lambda'}(j) e_\gamma^{-\lambda''}(k)}{\sqrt{\bar{M}_i \bar{M}_j \bar{M}_k}} e^{\pm i\mathbf{q}'\cdot\mathbf{r}_j} e^{-i\mathbf{q}''\cdot\mathbf{r}_k} \quad (S7)$$

$$V_{\lambda\lambda'\lambda''\lambda'''}^{(\pm\pm)} = \sum_{ijkl}\sum_{\alpha\beta\gamma\theta} \Phi_{ijkl}^{\alpha\beta\gamma\theta} \frac{e_\alpha^\lambda(i) e_\beta^{\pm\lambda'}(j) e_\gamma^{\pm\lambda''}(k) e_\gamma^{-\lambda'''}(l)}{\sqrt{\bar{M}_i \bar{M}_j \bar{M}_k \bar{M}_l}} e^{\pm i\mathbf{q}'\cdot\mathbf{r}_j} e^{\pm i\mathbf{q}''\cdot\mathbf{r}_k} e^{-i\mathbf{q}'''\cdot\mathbf{r}_l} \quad (S8)$$

Sn(II) has the lowest energy barrier along *a*- and *b*- axis of unit cell, indicating the weakest bonding with the lattice and the smallest linear restoring force.

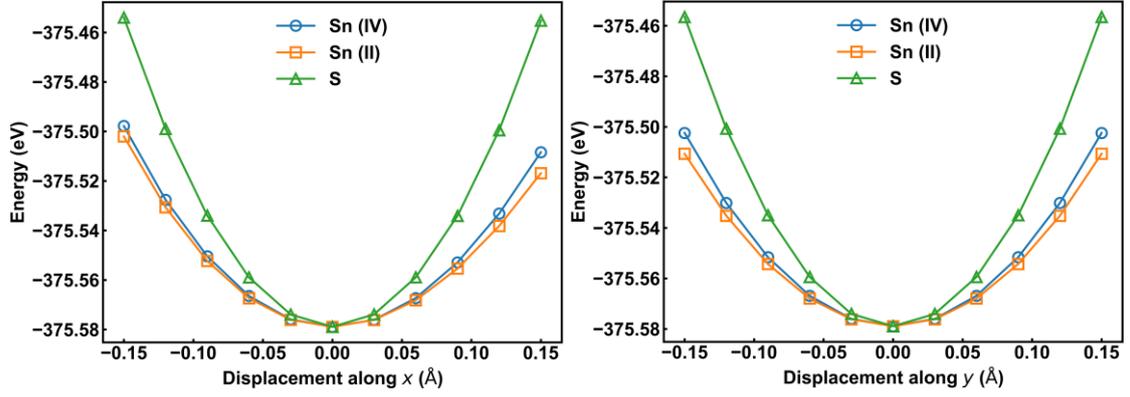

**Fig. S1.** Potential energy vs displacements along crystallographic *a*- and *b*- axis of unit cell.

To gain a deep understanding of bonding interactions, we present the projected COHP of $Sn_2S_3$ in **Fig. S2**. The active lone pair electrons are dominantly contributed by the s-orbital and exhibit anti-bonding interactions.

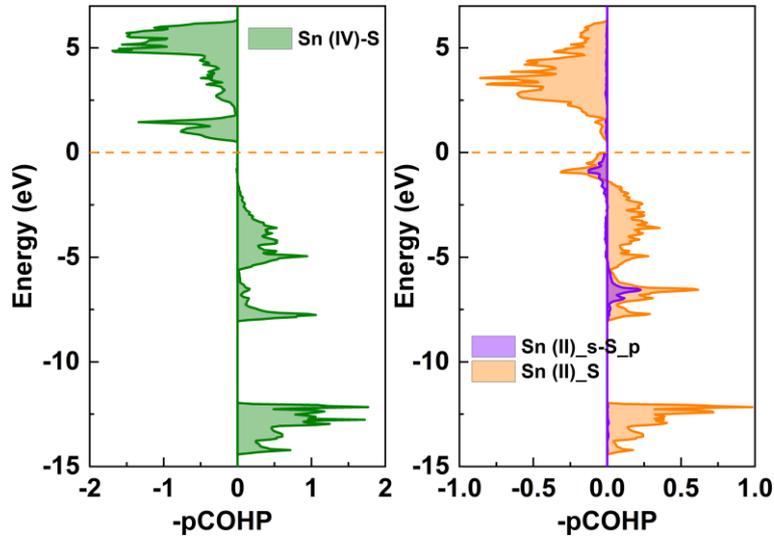

**Fig. S2.** -pCOHP of $Sn_2S_3$.

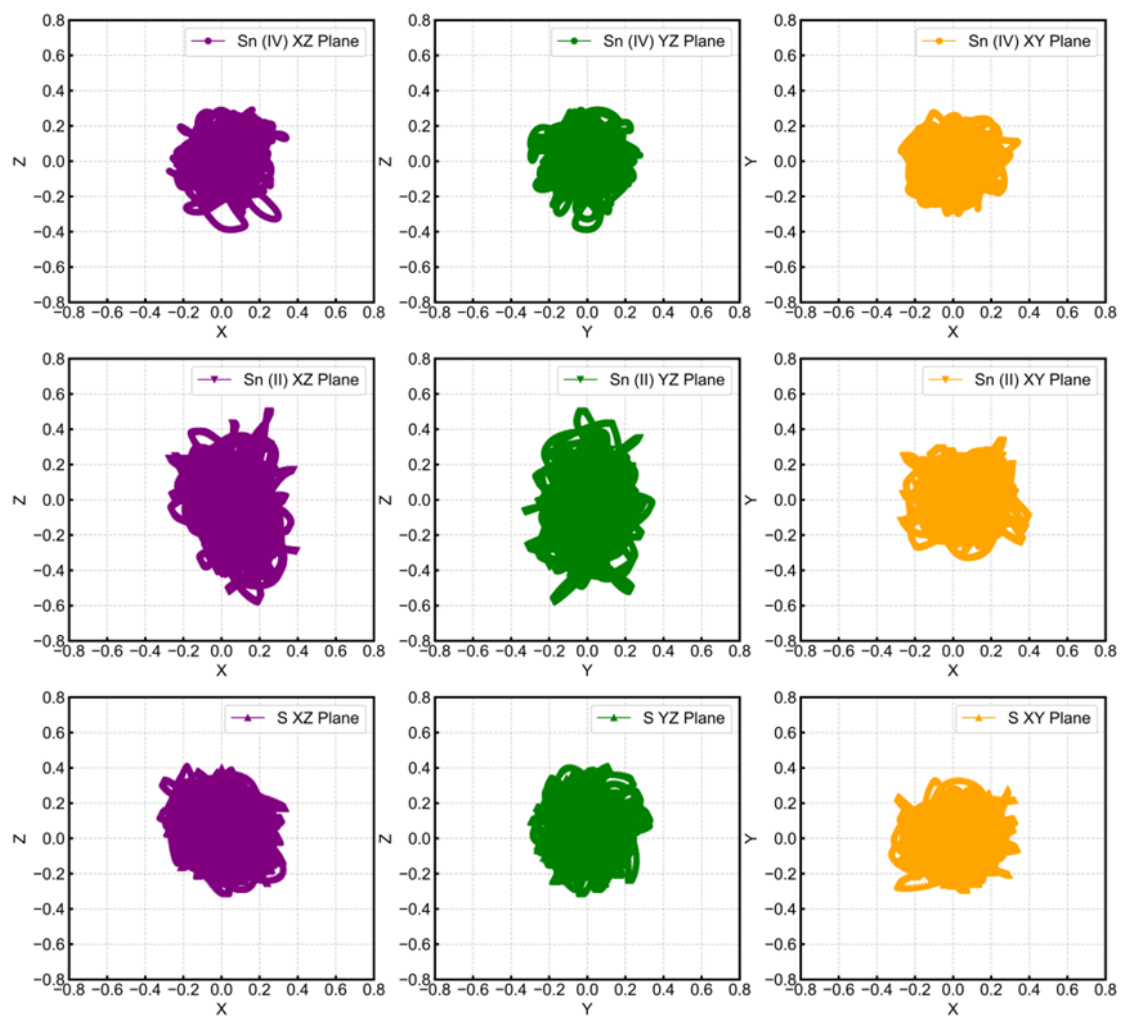

**Fig. S3.** Trajectories at 300 K of Sn(IV), Sn(II), and S Atoms in Sn$_2$S$_3$.

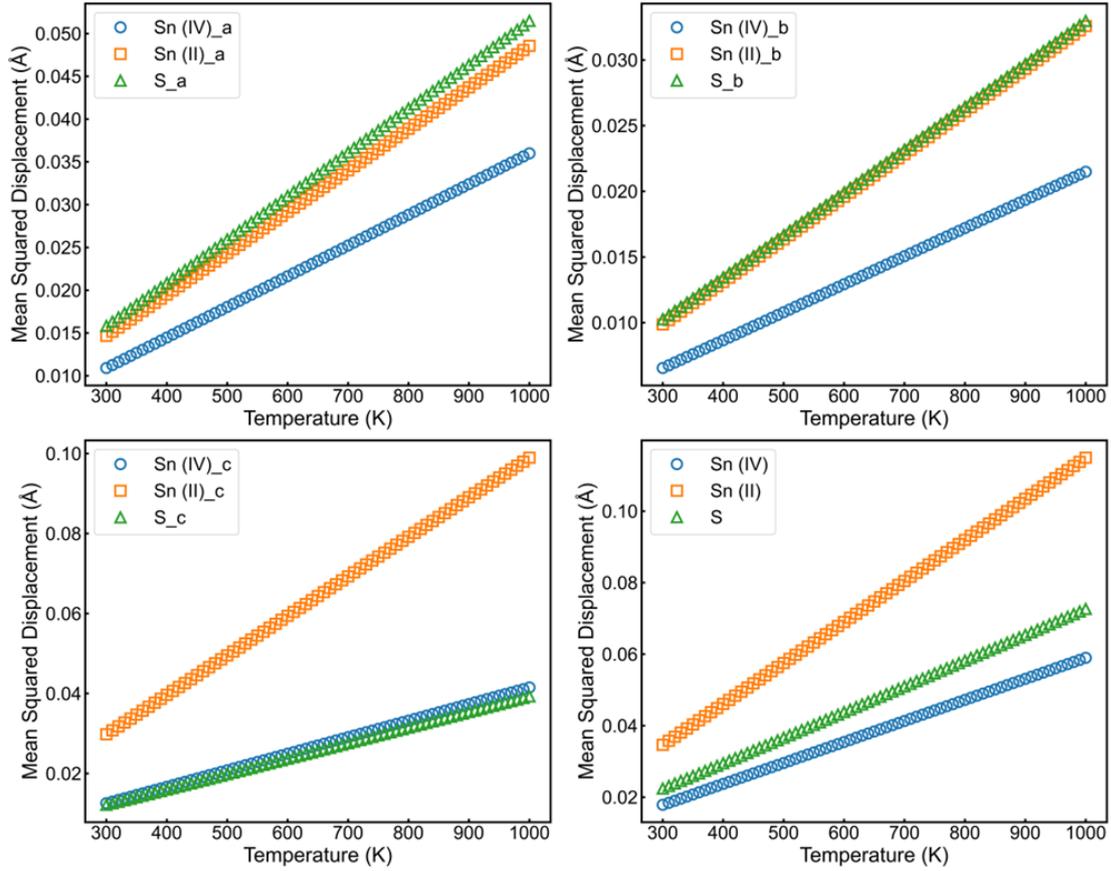

**Fig. S4.** Mean squared displacement (MSD) of Sn$_2$S$_3$ along the *a*, *b*, *c* directions and total MSD of Sn$_2$S$_3$.

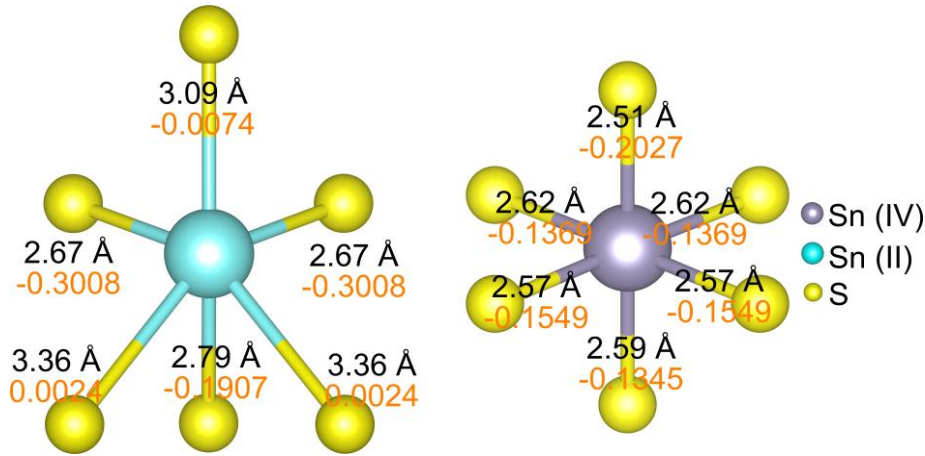

**Fig. S5.** Diagram of Sn(II) and Sn(IV) with their six nearest neighboring S atoms, respectively. where the black numbers represent bond lengths and the orange numbers represent normalized trace with interatomic bonding.

The thermal diffusivity of the Sn$_2$S$_3$ single crystal was characterized through cross-plane measurement from the top (105) surface to the bottom surface perpendicular to the (105) crystal plane. The uncertainties of thermal diffusion coefficient (*D*), the density ($\rho$), and specific heat capacity ($C_p$) are ~5%, ~5%, and ~15% in the measurements.

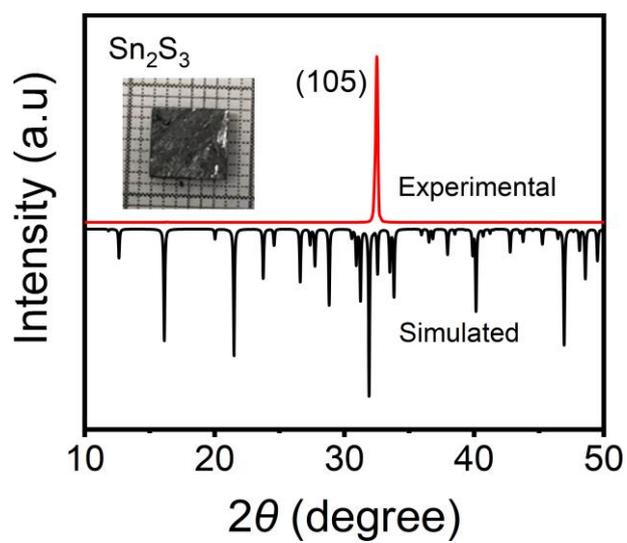

**Fig. S6.** (a) The PXRD (powder X-ray diffraction) pattern of the tested $Sn_2S_3$ single crystal sample, and the illustration shows the morphology of the sample.